# 5G Non-Public Network for Industrial IoT: Operation Models

Ahmad Rostami[a,1], Dhruvin Patel[b], Madhusudan Giyyarpuram[c], Finn Pedersen[a]
a) Ericsson AB, Sweden, b) Ericsson GmbH, Germany, c) Orange Lab, France

*Abstract*—5G non-public networks (NPNs) play a key role in enabling critical Industrial Internet of Things (IoT) applications in various vertical industries. Among other features, 5G NPNs enable novel operation models, where the roles and responsibilities for setting up and operating the network can be distributed among several stakeholders, i.e., among the public mobile network operators (MNOs), the industrial party who uses the 5G NPN services and 3rd parties. This results in many theoretically feasible operation models for 5G NPN, each with its own advantages and disadvantages. We investigate the resulting operation models and identify a set of nine prime models taking into account today's practical considerations. Additionally, we define a framework to qualitatively analyze the operation models and use it to evaluate and compare the identified operation models.

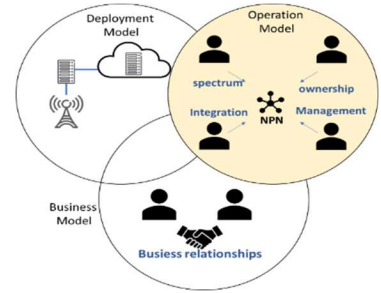

*Figure 1. Different dimensions of NPN realization. Personas represent different stakeholders in the ecosystem.*

## I. INTRODUCTION

The 5G mobile communication system introduces several advanced features [1], which have the potential to revolutionize the organization and operation of the machines as well as their interactions among themselves and with the human in a variety of vertical industries. Ultra-reliable and low latency communication (URLLC), network slicing, integrated edge computing, network exposure functions and modularity of the overall 5G system architecture are examples of these features. The features collectively provide the vertical industries with an innovation platform for optimizing the design and operation of their systems and for designing and implementing novel use cases. The application scenarios could range from Industrial Internet of Thing (IIoT) systems, e.g., in industrial automation, smart manufacturing, smart ports and utilities, to public safety, intelligent transportation systems (ITS) and healthcare.

For the IIoT systems, 5G provides the concept of non-public networks (NPNs), which is a new model for realizing and using cellular networks. The 3rd generation partnership project (3GPP) defines NPNs—also known as private cellular networks—as *intended for the sole use of a private entity such as an enterprise, and may be deployed in a variety of configurations, utilizing both virtual and physical elements* [2]. Thanks to the modularity and flexibility of the 5G system architecture, 5G NPNs can be realized in a variety of deployment models. A deployment model specifies the composition of 5G network functions, their placements across possible locations, e.g., local/on-premises vs remote/centralized, and the possible integration of these functions with those of a public mobile network. In every 5G NPN realization scenario, choosing the right deployment model will depend on the technical, operational, and business requirements of the scenario [3][4].

Besides the deployment models of 5G NPNs, which have been widely investigated over the last years, realization and operation of the NPNs require a good understanding of other less-explored dimensions, including operation models and business models (Figure 1). The operation models specify the allocation of roles, which are involved in the realization and operation of an NPN. For example, what stakeholder takes the role of spectrum provider or the role of the network operator.

As the NPN concept gains traction in the industry, national regulatory authorities in many countries across the world have started to develop new regulations to facilitate the operation of cellular NPNs, e.g., through direct licensing or leasing local spectrum to enterprises, who are willing to deploy and operate NPNs on their premises (e.g., see [5]). This together with the flexibility in NPN deployment model have the potential to disrupt the established ecosystem of mobile network operations. Specifically, today's public consumer-centric mobile network ecosystem realize a centralized operation model, where all the roles and responsibilities for realizing and operating mobile networks are with mobile network operators (MNOs). With the 5G NPN, there is for the first time the possibility to adopt novel decentralized operation models,

---

[1] The work of the author was partly conducted while he was associated with Robert Bosch GmbH, Germany.

where the roles and responsibilities for setting up and operating the network can be distributed over several stakeholders. For instance, consider a case where the NPN is designed and realized by a non-MNO 3rd party, the spectrum is provided by the enterprise, who is the user of the NPN, and the NPN is operated by an MNO. NPNs allow the adoption of centralized operation models run by the enterprise itself, or for example an MNO, as well as decentralized models where the stakeholders take different roles, as will be elaborated in this paper.

Our objective in this paper is to shed light on 5G NPN operation models, and in particular those models which are made possible by the introduction of 5G and local spectrum, as well as on the relation between the operation models and deployment models (Figure 1). Also, in this paper we identify and present a set of prime 5G NPN operation models for IIoT scenarios, discuss their inter-relations with the NPN deployment models, and qualitatively evaluate and compare the presented operation models. In our work, we do not investigate the NPN business models, which is also an important dimension of 5G NPN realization and operation (Figure 1). An example analysis of business models for 5G NPNs can be found in [6].

The remainder of this paper is organized as follows. In Section I, we review the major NPN deployment models as presented in [1] [2] [3]. In Section III, we define the elements of an operation model, explore the possible 5G NPN operation models and identify several prime models, without an assumption on the underlying deployment model. Then in Section IV, we investigate the interplay between the operation models and deployment models. In Section V, we focus on qualitatively analyzing the major NPN models identified in Section III from the NPN user perspective. Finally, in Section VI, we summarize our findings.

## II. 5G NPN DEPLOYMENT MODELS

The 5G system is based on the service-based architecture principles specified in 3GPP technical specification [1]. The 5G system architecture includes network functions performing various tasks to ensure data connectivity of the user equipment (UE) with the external data network. Figure 2(a) shows a simplified high-level 5G system architecture. The local/external packet network (e.g. industrial communication network) is connected to 5G system via user plane (UP). A user plane path is established between UE and the 5G core enabling bidirectional flow of the industrial application data. The user plane functions including the radio access network (RAN) are configured and controlled by the management and control plane (CP) of the 5G system. The RAN supports the UE mobility, access control and Quality of Service (QoS) realization over radio network. Furthermore, there is a group of management functions, which support the (re-)configuration of all network functions and corresponding services in the 5G system. Typically, the management functions are highly implementation-specific and independent from the chosen NPN deployment model and is therefore not considered for investigation in the paper.

From an architecture point of view, there are different possibilities through which a 5G NPN can be realized. Hence, it is of paramount importance to investigate the suitability of such options for a wide range of the IIoT use cases [2][1]. 3GPP distinguishes between two types of NPN deployments, namely Standalone NPN (SNPN) and Public network integrated NPN (PNI-NPN).

While the SNPN can be independent from public land mobile networks (PLMNs), the PNI-NPN relies on the network functions provided in PLMNs by the MNOs. Additionally, depending upon how the network functions are shared between NPN and PLMN, four deployment models are possible as illustrated in Figure 2(b) [3]. NPN1 assumes all the network functions of 5G network are on-premises and there is no connection towards a PLMN. SNPN can be realized also with sharing the RAN with the PLMN (i.e. NPN2). In NPN3, CP is hosted by PLMN and only the UP remain local on premise. In NPN4, the UP, CP and the RAN are all off premises.

## III. NPN OPERATION MODELS

In the previous section, deployment models are presented without discussing roles and responsibilities of various stakeholders in the operation of the 5G network. So far, only deployment models are described in 3GPP and in industrial fora such as NGMN [4] and 5G-ACIA [2]. Operation models are a way to take into account the roles of different stakeholders involved in operating an NPN [7][8][9]. An operation model specifies the assignment of roles to the stakeholders. The stakeholders and roles used in this paper represent a simplified but important subset of all stakeholders and roles in an ecosystem. This section considers the operation aspects of 5G networks and analyses various relevant scenarios. For this purpose, we consider the following stakeholders and roles. The stakeholders are:

*MNO* is the stakeholder who owns and manages a public land mobile network (PLMN). In view of the new ecosystem made possible by the 5G technology, MNOs also engage in value creation in vertical domains.

*Industrial Party* is the stakeholder who requests NPN services for performing a (group of) industrial task.

*3rd Party* is the stakeholder who provides equipment and/or services for deployment and management of NPN and cannot be categorized as MNO or industrial party.

And the roles are:

*NPN owner* is the role of owning the NPN infrastructure and includes both hardware and software components.

*Spectrum owner* is the role of having the right to transmit radio signals in a certain frequency band.

*NPN-integrator* is the role of setting up the NPN according to a chosen architecture making it ready to use

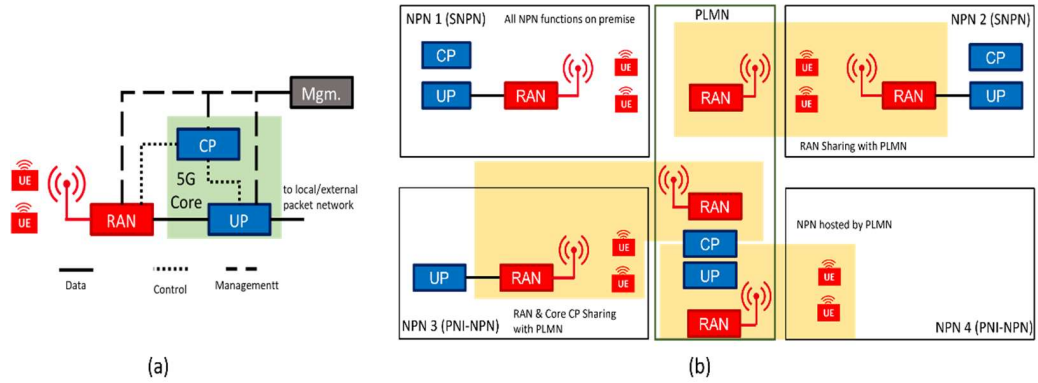

*Figure 2. (a)High-level 5G system architecture. UP: User Plane, CP: Control Plane, Mgm.: Management (b) NPN deployment model 1-4*

**NPN operator** is the role of operating and managing the NPN on a day-to-day basis. The NPN operator also offers NPN services, and as such the NPN service provider role can be assumed a sub-role of the NPN operator (as defined in [4]).
**NPN User** is the role who uses the services offered by the NPN for performing a (group of) industrial task.

For the sake of simplicity, the term ownership is used in a more generic sense, and we do not distinguish between direct or indirect ownership. For instance, in case of spectrum ownership we do not differentiate between the ownership as a licensee or through a leasing agreement with a licensee.

Except for the role of NPN user, which is exclusively assigned to the Industrial party, other roles can be taken by any of the three stakeholders. Therefore, if we put the NPN user aside, in theory 81 distinct operation models are identified (i.e., 3 (NPN Owner) × 3 (Spectrum Owner) × 3 (NPN Integrator) × 3 (NPN Operator). However, not all combinations would be meaningful in practice. Several factors such as business interests of stakeholders, or local regulations in different geographical areas make certain operational models more attractive than others and therefore more likely to materialize. Taking into account common practices and business considerations in ecosystems of today's mobile communication operation, a limited number of prime operation models can be identified. For example, providing the spectrum to a standalone NPN might not be business-wise justified for an MNO if it does not operate that NPN. That is, if an MNO acts as the spectrum owner, then it is likely that it also assumes the NPN operator role.

Accordingly, we have identified nine prime operation models as depicted in Table 1. In operation model 1 (OM1), the industrial party takes all the responsibilities for the NPN operation, i.e., the industrial party implements and integrates an NPN, obtains the spectrum for it, operates it and uses the corresponding NPN services without directly involving any other stakeholder. At the other end of the scale, we have operation model 9, where an MNO takes all the roles, and the industrial player, as the NPN user, relies on the services provided by the MNO. In between, there are options as represented by operation models 2 to 8, where the roles are assigned to two or more stakeholders.

IV. INTERPLAY BETWEEN OPERATIONAL MODEL AND DEPLOYMENT MODEL

In deriving the operation models in Table 1 there are no assumptions regarding the deployment models. In practice, however, the deployment and operation models are intertwined. Specifically, the choice of a deployment model can have an impact on the feasibility of an operation model and the other way around.

Table 2(a) illustrates the feasibility of all combinations of deployment models and operation models, where in total three distinct patterns can be observed. A key factor in the feasibility analysis of a combination is whether an MNO is the NPN owner or not. If not, then only NPN1 can be adopted, since sharing of resources with an MNO (NPN2-NPN4) does not make sense if those resources do not belong to the MNO. This is the case for seven out of nine operation models, i.e., models 1-3 and 5-8. In case of OM4, NPN1, NPN2 and NPN3 are feasible. NPN4 is not feasible in this case mainly because the industrial party is the spectrum owner, which prevents the integration of NPN into a PLMN with public spectrum. Finally, the operation model 9 features the highest level of flexibility, when it comes to the combination with deployment models. The reason is that in this model, MNO assumes all the roles and accordingly will have the flexibility to decide on the level of integration between NPN and the public network.

*Table 1. Prime Operation Models of 5G NPN*

|  | Owner | Spectrum | Integration | Operation |
|---|---|---|---|---|
| OM1 | Industrial Party | Industrial Party | Industrial Party | Industrial Party |
| OM2 | Industrial Party | Industrial Party | MNO | MNO |
| OM3 | Industrial party | Industrial Party | 3rd Party | 3rd Party |
| OM4 | MNO | Industrial Party | MNO | MNO |
| OM5 | 3rd Party | Industrial Party | MNO | MNO |
| OM6 | 3rd Party | Industrial Party | 3rd Party | 3rd Party |
| OM7 | 3rd Party | MNO | MNO | MNO |
| OM8 | 3rd Party | MNO | 3rd Party | MNO |
| OM9 | MNO | MNO | MNO | MNO |

*Table 2. Interrelation between operation and deployment models. A) Feasibility of combinations between deployment and operation models (OM). Green: combination is feasible. Red: combination is not feasible. B) Combinations of deployment models and shared operation roles. Green: combination is likely. Red: combination is not likely.*

(a)

|     | NPN1 | NPN2 | NPN3 | NPN4 |
|-----|------|------|------|------|
| OM 1 | green | red | red | red |
| OM 2 | green | red | red | red |
| OM 3 | green | red | red | red |
| OM 4 | green | green | green | green |
| OM 5 | green | red | red | red |
| OM 6 | green | red | red | red |
| OM 7 | green | red | red | red |
| OM 8 | green | red | red | red |
| OM 9 | green | red | red | red |

(b)

| | NPN Owner | NPN Operator | |
|---|---|---|---|
| | Shared btw. MNO & Ind. P | Shared btw. MNO & Ind.P | Shared btw. 3rd P & Ind. P |
| NPN 1 | red | green | green |
| NPN 2 | green | green | red |
| NPN 3 | green | green | red |
| NPN 4 | red | green | red |

In our analysis of the operation models and their interrelation with the deployment models (Table 1 and Table 2(a)) each role was exclusively assigned to one stakeholder, which in the rest of this paper we refer to as dedicated operation model. There are many situations, however, where it is desired or beneficial from technical, business, or operational perspective to share a role partially or fully between two stakeholders (In theory a role can be shared with more than two stakeholders, but it is unlikely in practice due to operational complexity). An operation model with a shared role in it is referred to as shared operation model. A prominent case for a shared operation model is where an industrial party—as the NPN user—would like to reduce the burden and complexity of the network operation and management on one hand, and on the other hand, have a certain level of visibility into the network management and handle simple management functions like end user activation and deactivation. In this case, it makes sense that the NPN operation role is shared between Industrial party and another stakeholder who takes NPN operation role, i.e., MNO or 3rd party. A similar sharing concept can also be applied to network ownership, where the network resources are shared between an MNO and an industrial party. Besides the roles of NPN owner and NPN operation, other roles, i.e., NPN spectrum owner, NPN integrator and NPN user, are not subject to sharing, mainly due to the nature of these roles.

There are also interrelations between NPN deployment models and possibilities for sharing of NPN operation roles. More specifically, not all NPN deployment models can be easily combined with shared operation models. Table 2(b) depicts the combinations. In NPN 1, where the NPN deployment is standalone and isolated from any MNO network, sharing of roles only make sense for the NPN operator role and between industrial party and a 3rd party who is tasked with the network operation and management. In this case, there will be little incentive for an MNO to take the responsibility of the NPN in a shared mode because the NPN is fully isolated from the PLMN. In NPN2 and NPN3 deployment models, where parts of the networking resources—i.e., RAN or Core—are shared between the industrial party and an MNO, it makes sense to also share the NPN ownership and NPN operation between the two stakeholders. Finally, in NPN4, where all resources are realized in a dedicated manner by an MNO, a shared operation model is only likely for NPN operation role, where network management and operation functions are partially shared between the two.

## V. EVALUATION OF THE OPERATION MODEL

In this section we evaluate the nine operation models. In our analysis we focus only on the dedicated operation models as presented in section III, since the number of the shared operation models can be excessively large depending on which part of a role (out of many) might be shared between two stakeholders. In our evaluations, the focus is put on the industrial party as the NPN user. That is, we evaluate each operation model against key criteria from the NPN user perspective. Our evaluations focus on technical and operational aspects of NPNs.

### A. Key Criteria

Below we first present the criteria that might affect the choice of operation models.

**NPN operation readiness** A 5G NPN has different control and management functions compared to other industrial communication technologies and will require a high level of competence to operate the network. NPN operation readiness considers the level of complexity that a stakeholder needs to deal with for operating a given NPN, which can vary depending on the type of the operation model considered where different stakeholders can assume different roles. Specifically, this criterion indicates how well a stakeholder needs to prepare to take up the NPN operation role.

**Service continuity** demonstrates the capability of ensuring connectivity when NPN devices move out of the coverage of an individual NPN. This is for example applicable for IIoT applications that require end-to-end connectivity across multiple NPNs and also in the PLMN.

**Privacy and security** demonstrate how an operation model fulfills the typically stringent security and privacy requirements of IIoT applications, which are critical for protecting sensitive operational data and control over the network infrastructure, e.g. for the security software updates. We assess the privacy and security jointly as a single criterion.

**Deployment flexibility** indicates the ability to set up and customize the network based on the needs of the IIoT applications. These needs may evolve over time and hence this flexibility is required over the lifecycle of the system/service. There are several possibilities in which an NPN network can be (re)configured to achieve certain objectives.

**Scalability** measures the ability to scale (e.g. expand the network capacity and/or coverage) the NPN network based on changing requirements or introduction of new IIoT use cases. These five criteria focus on technical and operational aspects of the NPN operation models. There are obviously other—mostly regulatory and business related—criteria, which influence the choice of an operation model for IIoT scenarios. For instance, some NPN users may have multiple sites spread across the globe and may desire to make common deployment and operational choices. Global applicability is the criterion that captures this capability. The choice of the operation model based on global applicability highly depends on the regulatory aspects such as the availability of the spectrum options in selected countries of NPN deployment. A detailed investigation of local spectrum options and their relations to the NPN operation models is a very broad topic and goes beyond the scope of this paper. For example, an industrial party can only be a spectrum owner if it can be a licensee or a lessee of the spectrum. If the regulation allows, it is possible to utilize both private industry spectrum and public spectrum for achieving specific purposes (In this case there are two separate Spectrum Owners, the role is not shared). For instance, one can combine private and public spectrum to increase the available bandwidth. Also, it is possible that an MNO can offer an NPN solution operating in private spectrum instead of MNO licensed public spectrum [10].

Another crucial criterion is the total cost of operating NPN. While several of the above KPIs have an impact on costs, a complete evaluation of the costs often requires information on commercial offers, which are not publicly available.

B. *Analysis of operation model*

For qualitatively evaluating the operation models against the criteria defined above three qualifiers are used: *High, Medium, and Low*. *High* means that the corresponding criterion is very well supported under the considered NPN operation model. *Low* is used to indicate that the criterion is either not supported or supported only with significant adaptations. Finally, *Medium* is used as a level between *High* and *Low*. As we will see below, the evaluation of operation models will in many cases depend also on the adopted deployment model. Accordingly, *Table 3* presents three sets of evaluation results: the results for the case that NPN1 is adopted (applicable to all operation models), the results for the cases that NPN2 or NPN3 are adopted (applicable to OM4 and OM9), and finally those for scenarios with NPN4 (applicable only to OM9).

In evaluating the NPN operation readiness, we argue that it only depends on who operates an NPN. If an NPN is operated directly by an Industrial party (i.e., OM1), then the initial operation readiness would be Low, since the operation of a cellular network requires a certain level of competence, which is usually not available within an OT enterprise and needs to be build up. In contrast, the operation readiness level is evaluated as High in all other cases (OM2-OM9) since the operation is done by an experienced MNO or 3$^{rd}$ party.

Similar to the NPN operation readiness, the service continuity of operation models also depends on who operates the NPN in question. Here, if the NPN operator is not an MNO (i.e., OM1, OM3 and OM6), then the service continuity will be considered low, since it requires a separate agreement between the NPN operator and an MNO, which needs to be negotiated. On the other hand, for other models, where an MNO operates the NPN, the service continuity can be evaluated between Medium to High depending on the adopted deployment model. Specifically, offering service continuity would be much easier (High) if the NPN is highly or fully integrated in the PLMN (NPN 4 and NPN 3) and it will be less straight forward (Medium) if the NPN is stand-alone (NPN 1) or only partially integrated into the PLMN.

The privacy and security of operation models depend on who will act as the NPN owner and NPN operator, since these might have access to the user data and metadata (e.g., NPN control and management data). There are established methods that can be applied to ensure the requested security and privacy of the user data and metadata in all the operation models in Table 1. Examples of such methods include application-level encryption, service level agreements (SLA) and network slicing [11]. Having said that, the security and privacy KPI for various operation models can still be evaluated and ranked based on a) the potential visibility of metadata to actors other than the Industrial party and b) the degree of control that the Industrial party is able to exercise on operations. Accordingly, we evaluate the security and privacy as High for the OM1, where only the industrial party has visibility to the metadata, and it has full control over the adopted security and privacy measures. All other operation models (OM2-OM9) are evaluated as Medium or Low in Table 3 depending on if one or two other stakeholders (besides the industrial party) are involved in ownership and operation of the NPN, respectively. Obviously, a higher number of stakeholders might increase the potential security and privacy risks (the so-called attack surface) and make the control over security measures more complicated.

The deployment flexibility depends on the allocation of NPN owner and operator roles. In particular, the industrial party will enjoy the highest level of flexibility in customizing the NPN deployment if both the NPN owner and operator roles are assigned to it (OM1). In other cases, the industrial party need to rely—at least partially—on 3$^{rd}$ party or MNO capabilities for NPN customizations. In these cases, if the industrial party is still the owner (OM2 and OM3), we evaluate it as Medium, since it will still have good leverage to customize the NPN. In other cases, i.e. OM4, OM5 and OM7-OM9, the flexibility is considered Low. The only exception here is the case, where a 3$^{rd}$ party is both the NPN owner and operator. This case we evaluate also as Medium, since a 3$^{rd}$ party might provide comparatively a better support for customization than an MNO.

For the scalability of the operation models, the adopted NPN deployment model plays a key role. Specifically, the more an NPN is integrated into the PLMN, the higher its scalability will be, because MNOs usually have great amounts of resources at hand to scale up a PNI-NPN if needed. Accordingly, we evaluate the scalability of operation models adopting NPN1 with Low, those adopting NPN2-NPN3 with Medium and the ones adopting NPN4 with High.

Another important aspect of the NPN operation models analysis, which is not directly obvious from Table 3, is the impact of direct spectrum allocation to the industrial party (the so-called local spectrum). From an industrial party's point of view the interest in local spectrum could be due to independence from an MNO (i.e., OM1 will not be possible without the local spectrum) and/or having dedicated spectrum for the IIoT use cases. In practice, an MNO has all the means to fulfill the latter requirement, e.g., through adopting network slicing. Therefore, as long as the RAN equipment supports the frequency bands used for local spectrum, the functioning of the system is not affected by the fact that local spectrum is used.

In Summary, there is no single model that optimizes all the criteria. It also illustrates that an NPN user has to make trade-offs when choosing the right operation model. As an example, *Table 3* demonstrates a clear trade-off between NPN operation readiness on one hand and privacy and security as well as deployment flexibility on the other hand, such that not all these criteria can be optimized at the same time. Therefore, the right operation model varies depending on how different criteria are prioritized for a certain IIoT scenario. For instance, assuming the NPN1 deployment model, an NPN user might choose:
- OM1 to have High for security and privacy,
- OM2, OM3 or OM6 to get at least Medium for security and privacy, as well as for operation readiness and deployment flexibility.

Table 3. Operation models Analysis. For results indicated with a, b, and c the assumption is that deployment models NPN1, NPN2/NPN3 and NPN4 are adopted, respectively. If not indicated, results apply to all NPN1-NPN4 (if applicable according to Table 2(a))

| | NPN Operation Readiness | Service Continuity | Privacy & Security | Deployment Flexibility | Scalability |
|---|---|---|---|---|---|
| OM 1 (NPN1) | Low | Low | High | High | Low |
| OM 2 (NPN1) | High | Medium | Medium | Medium | Low |
| OM 3 (NPN1) | High | Low | Medium | Medium | Low |
| OM 4 (NPN1-3) | High | Medium | Medium | Low | Low[a] (Medium[b]) |
| OM 5 (NPN1) | High | Medium | Low | Low | Low |
| OM 6 (NPN1) | High | Low | Medium | Medium | Low |
| OM 7 (NPN1) | High | Medium | Low | Low | Low |
| OM 8 (NPN1) | High | Medium | Low | Low | Low |
| OM 9 (NPN1-4) | High | Medium[a,b] (High[c]) | Medium | Low | Low[a] (Medium[b]/High[c]) |

## VI. Conclusion

We identified nine prime operation models of 5G NPN for IIoT scenarios and elaborated on how the models differ from each other depending on distribution of selected roles and responsibilities among selected stakeholders. Also, we proposed a framework for a systematic analysis of NPN operation models, including a set of criteria from an NPN user perspective to qualitatively analyze and compare the models. Our analysis highlights the trade-offs that are involved in the selection of the right NPN operation model. Each NPN user has to apply its own weights and priorities in order to derive preferable options. While our analysis focused on the technical and operational aspects of the NPN operation, the final choice of the NPN operation model will depend also on other regulatory and business-related criteria such as spectrum allocation policies, cost, liability protection, lock-in protection, and business models.

## Acknowledgment

Part of this work has been performed in the framework of the H2020 project 5G-SMART co-funded by the EU. Authors would like to acknowledge the contributions of their colleagues from 5G-SMART although the views expressed are those of the authors and do not necessarily represent the views of the 5G-SMART project.

## References

[1] 3GPP technical specification TS 23.501, "System architecture for the 5G System (5GS)", Release 16, August, 2020.

[2] 3GPP technical specification TS 22.261, "service requirements for the 5G system," 2020.

[3] "5G for automation in industry," 5G-ACIA Whitepaper, March 2019. [Online]. Available: https://www.5g-acia.org/index.php?id=6960

[4] NGMN, 5G E2E technology to support verticals URLLC requirement, October, 2019.

[5] German Federal Ministry for Economic Affairs and Energy, "Guidelines for 5G Campus Networks – Orientation for Small and Medium-Sized Businesses," April 2020. [Online]. Available: https://www.bmwi.de/Redaktion/EN/Publikationen/Digitale-Welt/guidelines-for-5g-campus-networks-orientation-for-small-and-medium-sized-businesses.pdf?__blob=publicationFile&v=2

[6] Ahokangas et al., "Business Models for Local 5G Micro Operators", IEEE Transactions on cognitive communications and networks, Vol 5, No 3, September 2019, pp 730-740.

[7] A. Rostami, "Private 5G Networks for Vertical Industries: Deployment and Operation Models," 2019 IEEE 2nd 5G World Forum (5GWF), Dresden, Germany, 2019, pp. 433-439, doi: 10.1109/5GWF.2019.8911687.

[8] 5G-SMART, Deliverable 5.2, "5G network architecture options and assessments", November 2020.

[9] 3GPP technical report TR 28.807, "Study on management of Non-Public Networks (NPN)," Release 16, 2020.

[10] Mats Norin, *et al.*, "5G spectrum for local industrial networks," Ericsson Whitepaper, April 2021. [Online]. Available: https://www.ericsson.com/en/reports-and-papers/white-papers/5g-spectrum-for-local-industrial-networks

[11] "Security Aspects of 5G for Industrial Networks," 5G-ACIA Whitepaper. [Online]. Available: Security Aspects of 5G for Industrial Networks - 5G-ACIA (5g-acia.org)